\documentclass[12pt]{article}
\usepackage{graphicx,amssymb}

\makeatletter
\renewcommand\section{\@startsection {section}{1}{\z@}%
                                 {-3.5ex \@plus -1ex \@minus -.2ex}
                                   {2.3ex \@plus.2ex}%
                                   {\normalfont\large\bfseries}}
\renewcommand\subsection{\@startsection{subsection}{2}{\z@}%
                                   {-3.25ex\@plus -1ex \@minus -.2ex}%
                                     {1.5ex \@plus .2ex}%
                                     {\normalfont\bfseries}}
\renewcommand\subsubsection{\@startsection{subsubsection}{3}{\z@}%
                                   {-3.25ex\@plus -1ex \@minus -.2ex}%
                                     {1.5ex \@plus .2ex}%
                                     {\normalfont\itshape}}
\makeatother

\newcounter{multieqs}

\newcommand{\bq}{\begin{equation}}
\newcommand{\fq}{\end{equation}}
\newcommand{\bqr}{\begin{eqnarray}}
\newcommand{\fqr}{\end{eqnarray}}
\newcommand{\non}{\nonumber \\}

\def\alp{\alpha}   \def\bet{\beta}    \def\gam{\gamma}
\def\del{\delta}    
    \def\th{\theta}     
       \def\lam{\lambda} 
 \def\sig{\sigma}   
  \def\ome{\omega}

      \def\The{\Theta}
\def\Lam{\Lambda}

 \def\cH{{\cal H}} 
  
 \def\cN{{\cal N}} \def\cO{{\cal O}}

 \def\cZ{{\cal Z}}

\def\pa{\partial}

\def\inv{^{-1}} 

\def\pr{^{\prime}}

\def\rar{\rightarrow}

\newcommand{\tr}{\mbox{Tr}}

\def\hlf{\frac{1}{2}}
\def\ove#1{\frac{1}{#1}}

\def\RR{\mathbb{R}}

\def\m{{\rm m}}

\begin{document}
\thispagestyle{empty}
\begin{flushright}
\parbox[t]{2in}{CU-TP-1067\\
UPR-1009-T\\
MAD-TH-02-1\\
  hep-th/0208163}  
\end{flushright}

\vspace*{0.15in}

\begin{center}
{\Large \bf 
On the Hagedorn Behaviour of PP-wave Strings\\ and \\[0.12in]
$\cN=4$
  SYM Theory at Finite $R$-Charge Density}

\vspace*{0.5in} 
{Brian R. Greene${}^{1,2}$,~Koenraad Schalm${}^{3}$,~
and Gary Shiu${}^{4,5}$}\\[.3in]
{\em ${}^1$ Institute for Strings, Cosmology and Astroparticle Physics
\\ 
${}^2$ Department of Mathematics \\
${}^3$ Department of Physics  \\
Columbia University \\
New York, NY 10027}\\[0.2in]
{\em ${}^4$Department of Physics and Astronomy\\
University of Pennsylvania\\
Philadelphia, PA 19104}\\[0.2in]
{\em 
${}^5$ Department of Physics\\
  University of Wisconsin\\ 
Madison, WI 53706}\\[0.5in]
\end{center}

\begin{center}
{\bf
Abstract}
\end{center}
We discuss the high temperature behaviour of IIB strings in the
maximally symmetric plane wave background, and show
that there is a Hagedorn temperature.  We discuss 
the map between strings in the
pp-wave background and the dual superconformal field theory in the
thermal 
domain.
The Hagedorn bound
describes a curve in the 
$R$-charge chemical potential versus temperature
phase diagram of the dual Yang-Mills theory and the
theory manifestly exists on both sides. 
Using a recent observation of Brower, Lowe,
and Tan,
we update our earlier calculation to reflect that the pp-wave
string exists on both sides of the Hagedorn bound as well.

\vfill

\hrulefill\hspace*{4in}

{\footnotesize
${}^1$ Email addresses: greene@phys.columbia.edu,
kschalm@phys.columbia.edu, shiu@physics.upenn.edu, shiu@physics.wisc.edu.
}

\newpage

\section{Introduction}

For many decades now, string theory has held the promise of answering
long  
standing mysteries that require a consistent theory of quantum
gravity. The two  
most pressing areas are those of black hole physics and cosmology,
both of  
which, in standard treatments, are afflicted by spacetime
singularities that  
conclusively demonstrate the need for a more robust theoretical
framework.   
During the last few years, significant progress has been made in
understanding  
aspects of black hole physics in the context of string theory
\cite{BH}, although  
the proper resolution of the Schwarzschild black hole singularity
remains  
elusive. More recently, there has also been a renewed surge of
activity in  
applying string theory to issues relevant to cosmology through the
advent of  
brane-world models \cite{HW,ADD,ShiuTye,Ovrut,RS}, the focus on de
Sitter spacetime \cite{Strominger,Witten,vijay}, the  
application of holographic ideas
\cite{FischlerSusskind,EastherLowe,BanksFischler,LvdSL}, and the
controversial possibility that  
string theory (and quantum gravity more generally) might leave an
observable  
imprint in the cosmic microwave background radiation
\cite{CGS,EGKS,Brandenberger,Niemeyer,KKLS,ShiuWasserman,Danielsson}. 

It has long been recognized that one requisite element of string
cosmology is  
an understanding of string theory at finite temperature, and much work
along  
these lines has been undertaken
\cite{thermal,Alvarez1,Alvarez:1987sj,Bowick:1989us,Atick:1988si}.
These works have also demonstrated that 
finite temperature studies have the capacity to shed much light on
critical,  
foundational issues of string theory itself. As a prime example,
compared with  
ordinary quantum field theory, string theory is exceedingly well
behaved in the  
ultraviolet, a property largely due to the exponential growth of
states as a  
function of energy. And since the earliest days of string theory it
has been  
recognized that such a growth of states leads to a Hagedorn
temperature  --- a  
temperature above which the statistical partition function diverges
because the  
exponential Boltzmann suppression is overcome by the exponential
density of  
states. The existence of a Hagedorn temperature is particularly
tantalizing  
because a long-standing, critical question has been: what are the true,
fundamental  
degrees of freedom in string theory? While strings provide a natural
and  
perturbatively useful set of constituents, the existence of a Hagedorn
temperature is an indication that
strings --- 
somewhat like low energy/temperature hadrons in quantum
chromodynamics --- may not  
be the elementary degrees of freedom of string theory. Perhaps, many
authors  
have speculated, just as quark and gluons emerge as the basic
ingredients of  
quantum chromodynamics at high enough temperatures (and energies), the
true  
degrees of freedom of string theory may also emerge at sufficiently
high  
temperatures. In fact, if the specific heat at the Hagedorn
temperature is  
finite, this indicates that the Hagedorn temperature is not a 
limiting    
temperature, but, instead, demarcates a phase transition --- one that
might well  
signal the appearance of the true, elementary degrees of freedom of
string  
theory.

Over the years there have been a number of studies of the Hagedorn
temperature  
in string theory; most 
have focused on string propagation in flat 
backgrounds since the calculation of the density of states requires an
exactly  
soluble model. For example,  in ref. \cite{Alvarez:1987sj} open
bosonic, open and  
closed supersymmetric, and heterotic string theories were studied in
flat  
noncompact space and it was found that the Hagedorn temperature is a
limiting  
temperature for the open string models but not for the closed string
models. In  
\cite{Bowick:1989us}, closed strings were studied on a flat compact
manifold ($T^9 
$) and it was found that the Hagedorn temperature is a limiting
temperature.  
More recently, \cite{abel,dienes} have studied D-brane thermodynamics
and found, for example,  
that D$p$-branes for $p < 5$ possess a non-limiting Hagedorn
temperature and  
hence likely have a high temperature Hagedorn 
phase.

In spite of many studies a precise
understanding  
of the properties that a Hagedorn phase in string theory would
exhibit, is still lacking.\footnote{Interesting recent progress
  on this question is reported in
  \cite{Barbon:2001di}.} A  
number of the works cited did reveal one likely qualitative feature of
strings  
near the Hagedorn temperature --- namely, the coalescing of energy into
one long  
string, but very little else is known. In light of the work of
\cite{Atick:1988si} 
this may not be particularly surprising. Through indirect reasoning,
these  
authors showed that the Hagedorn phase is characterized by a drastic
reduction  
in the number of degrees of freedom (the free energy behaves as $T^2$
instead  
of $T^D$) and hence it may well be a highly nontrivial task to
identify and  
rigorously describe the correct high temperature constituents. 

In this paper, we undertake a further study of high temperature string
theory  
in light of two recent and related advances. The first advance is the
recent  
realization that string propagation in a maximally supersymmetric 
pp-wave background provides a
non-flat, 
exactly soluble string theory
\cite{Metsaev:2001bj,Russo:2002rq}. It is natural,
then, to consider thermal  
string theory in such a background since we retain the necessary
analytical  
control to make explicit calculations even though the space is not
flat. The  
second advance is the realization that because of the AdS/CFT
correspondence,  
pp-wave string theory is dual to a subsector of a particular Yang-Mills 
theory \cite{Berenstein:2002jq}. 
This holds out the promise of a new, dual description of the Hagedorn 
properties of string theory (in a pp-wave background) and hence the
possibility  
of gaining insight inaccessible by more direct means.
    In what follows, through direct calculation, we will partially
    realize  
these expectations. 

In particular, in section 2 we will lay out the
basic  
formalism for explicit calculations of thermal strings in a pp-wave
background.  
In section 3 we will discuss the high temperature behaviour of strings
in a pp-wave background, 
and will show that, indeed, there is a Hagedorn
temperature.  
Close examination of the thermal properties of strings near the
Hagedorn  
temperature will reveal it to be a limiting
temperature.\footnote{Recent work of \cite{blt} modifies some of the 
  conclusions presented here. Extending our results (by analyzing full
  thermodynamic integrals rather than just their integrands as done here) \cite{blt} shows
  that the Hagedorn temperature rather signals a phase transition.}  
This is
counter to the intuition that at high energies 
the pp-wave string only differs mildly
from the Minkowski string. 
Therefore, it appears that we cannot gain much insight into
the nature of a Hagedorn phase (and the fundamental degrees of
freedom of string theory at high temperature) 
by studying strings in the pp-wave background.     
However, in section 4
we study  
the dual description of these results in ${\cal N}= 4$ $SU(N)$
super-Yang-Mills  
theory, and obtain a somewhat unexpected result. 
Extending the map between pp-waves and the dual
conformal  
field theory given in \cite{Berenstein:2002jq} to the thermal domain,
we find that the string Hagedorn bound is a function of the
super-Yang-Mills temperature and $R$-charge chemical potential.  
Surprisingly,   
the 
known regimes of the dual Yang-Mills
theory at finite temperature and $R$-charge density 
lie beyond the Hagedorn bound, and hence are inaccessible 
to 
the thermal 
pp-wave string. This suggests that 
the
limiting Hagedorn
temperature may not be absolute, and that there exists a physical
regime on the other side of the 
bound.\footnote{In fact, it turns out that this conjecture is true
  \cite{blt}; see note 
added in proof for details.} To further motivate
this conjecture, we note that thermodynamics in the
context of quantum gravity is
notoriously subtle, and it is questionable whether the ideal gas
approximation --- the approximation 
we use in our calculations --- is a valid one, as in this approximation one 
effectively tunes the
coupling to
zero. Perhaps beyond the ideal gas
approximation, the Hagedorn temperature ceases to be limiting and
the pp-wave string probes the full phase diagram of the SYM theory. 
In section 5 we offer our conclusions
and  
directions for further work.

While completing this paper, the preprint
\cite{PandoZayasVaman} appeared, 
which, in the revised version, contains
the same results on
the Hagedorn temperature and limiting behaviour of the IIB pp-wave
string as we find in our sections 2 and 3.

\section{Thermodynamics of pp-wave strings}
\setcounter{equation}{0}

\subsection{Partition function in a pp-wave background}

Our understanding of statistical mechanics, as of quantum field
theory, is ultimately based on a Hamiltonian formulation of the
theory. As such 
statistical mechanics suffers from the same complications
as quantum field theory when one considers the system in a space-time
background which is not flat. Only when the background space-time has
a timelike Killing vector, can we at present
make sense of the notion of a Hamiltonian as the generator of
time-evolution and construct an associated quantum field theory or
statistical partition function.\footnote{Thermodynamics
is strictly defined in the infinite volume limit, which for string
theory, or any quantum 
theory of gravity, does not
exist due to Jeans' instability. For sufficiently small string
coupling this problem may be ignored, which is the point of view we
shall take. We also assume that use of 
the canonical ensemble is valid.}

A first requirement, therefore, is to analyze the
symmetries of the pp-wave background. These have been extensively
discussed in \cite{blau,Michelson:2002wa}. The
maximally symmetric pp-wave background for the IIB string, to
which we will limit our attention, has the metric 
\begin{eqnarray}
  \label{eq:1}
  ds^2 = -dx^+dx^--\mu^2z_i^2(dx^+)^2+dz_i^2 ~~;~~~~i=1,\ldots,8
\end{eqnarray}
with the curvature supported by a constant five-form flux 
\begin{eqnarray}
  \label{eq:2}
  F_{+1234} = \frac{\mu}{g_s}.
\end{eqnarray}
The symmetries of the metric are two translational isometries in
the $dx^{\pm}$ directions, a rotational $SO(8)$ symmetry, broken to
$SO(4) \times SO(4)$ by the presence of the five-form flux, and eight
Lorentz-boost-like rotations in the $z^i,~x^-$ plane. 
In addition there is
the Lorentz boost $x^+ \rar \alp x^+, x^- \rar
x^-/\alp$, which is only a  
symmetry when accompanied by the rescaling $\mu
\rar \mu/\alp$. 

Consider now an arbitrary linear combination of
the translational isometries 
\begin{eqnarray}
  \label{eq:5}
  \xi(a,b) = a\frac{\partial}{\partial x^+} + b
  \frac{\partial}{\partial x^-}~.
\end{eqnarray}
We may
compute its nature, 
and construct a time-like Killing vector,
according to which we quantize the system. 
The length squared of
$\xi(a,b)$ is
\begin{eqnarray}
  \label{eq:3}
  ||\xi||^2
  =-ab-\mu^2z_i^2a^2 ~;
\end{eqnarray}
hence $\xi(a,b)$ is timelike if $-a(b+\mu^2z_i^2a) <0$. 
For instance, the naive choice 
$a=b=1$, in which case the Killing vector is simply
$\partial/\partial t$, ($x^{\pm} = t\pm r$), obeys this
inequality. The dependence of the length of $\xi(a,b)$ on the position $z$
simply reflects the redshift of temperature  
for various classes of
observers.\footnote{
These are hypothetical non-geodesic 
observers located at fixed $z$, which are the
natural ones to consider in the formal construction of the statistical
partition function as a weighted distribution of Hamiltonian
eigenvalues. Physical observers would follow geodesics instead.} 
We will choose both $a > 0$ and $b > 0$ to ensure that the
Killing vector is everywhere timelike. Quantizing the pp-wave system
with such a $\xi(a,b)$ as
Hamiltonian, the 
pp-wave partition function is formally
\begin{eqnarray}
  \label{eq:4}
  \cZ(a,b;\mu) = \tr_{\cH} \,  e^{-bp_--ap_+}
\end{eqnarray}
with the standard relation $p_{\pm} = -i\partial/\pa x^{\pm}$. We have
denoted the implicit dependence on the curvature of the pp-wave
background by the variable $\mu$. 

The symmetry of the theory under Lorentz scaling plus boost implies
that the partition function (\ref{eq:4}) only depends on dimensionless
combinations of $a,~b$ and $\mu$. This still means that the partition
function is
a function of {\em two} 
variables,
rather than a single temperature.  
The pp-wave geometry is not
isotropic and these two parameters correspond geometrically to the 
length of the time-like 
Killing vector and its angle with respect to an arbitrarily chosen
fiducial tangent vector field. 
The first selects the family of observers and the
second the temperature of the heat-bath this family measures. In flat
space the first variable, the angle, is irrelevant, as isotropy
assures that all families of observers are equivalent. The
identical expression for the partition function eq. (\ref{eq:4}) in
flat space, where $\mu=0$, can be Lorentz-boosted to a standard
one-parameter partition function with temperature $T\inv=\sqrt{ab}$. 
The Lorentz-boost is a symmetry of the flat space, 
$\mu=0$ theory, and the flat space 
partition function computed before or after the boost will give the
same answer. The pp-wave geometry does not possess this
symmetry, and the partition function (\ref{eq:4}) will therefore
be a two-parameter function of $a$ and $b$. In section
\ref{sec:impl-cft-dual} we will see that the dependence on 
the second parameter 
corresponds to the inclusion of a chemical potential in the dual
CFT partition function.\footnote{We thank Jan de Boer for pointing
  this out to us.} 

Let us conclude with a final comment. The pp-wave metric
is not static. Hence straight Euclideanization of the
spacetime will cause the metric to be complex. This
does not mean that we cannot discuss thermal physics. As emphasized at
the beginning of this section, to construct a partition function, 
the metric only needs to possess a
timelike Killing vector to serve as Hamiltonian --- the
spacetime must be stationary. In that case the concept of thermal
equilibrium can make sense. If spacetime is also static,
constructing a path integral from the statistical partition function
will be equivalent to compactifying Euclidean time. If the spacetime
is not static, one has
to resort back to the Hamiltonian approach, as we do here.  
 
\subsection{Single and multi-string partition functions from the lightcone}

As is well known, the partition function of an ideal gas of IIB pp-wave
strings is a weighted product of the partition function of a single
IIB 
pp-wave string (see e.g.
\cite{Alvarez1,Alvarez:1987sj,Bowick:1989us}) 
\begin{eqnarray}
  \label{eq:6}
  \ln \cZ(a,b;\mu) = \sum_{r=1}^{\infty} \ove{r} \left
  ( Z_1^B(ar,br;\mu) -(-1)^r Z_1^F(ar,br;\mu) \right)~.
\end{eqnarray}
This relation holds for a system of weakly interacting strings
immersed in a heat bath of temperature $T\inv(z) =\sqrt{ab+\mu^2z^2a^2}$. 
By definition the bosonic and fermionic components of the single
string partition function are
\begin{eqnarray}
  \label{eq:33}
  Z_1^B(\beta) &=& \hlf \tr(1+(-1)^{F_{st}})e^{-\beta H}~,\\
Z_1^F(\beta) &=& \hlf \tr(1-(-1)^{F_{st}})e^{-\beta H}~,
\end{eqnarray}
where $F_{st}$ is the spacetime fermion number operator. The partition
function is more conveniently expressed in terms of the building
blocks
\begin{eqnarray}
  \label{eq:34}
  Z_1^{Thermal}(\beta) = \tr e^{-\bet H} ~~,~~~~ 
Z_1^{WI}(\beta)=\tr (-1)^{F_{st}}
  e^{-\bet H}~.
\end{eqnarray}
For a spacetime supersymmetric string, as we have in mind, the Witten
index $Z_1^{WI}(\bet)$ counts the number of bosonic minus fermionic
zero energy groundstates. 
This is a finite number (it equals unity, see
\cite{Takayanagi:2002pi}) and
gives only a small temperature independent 
correction to the free energy. As we will be interested in the high
temperature behaviour of the pp-wave string, we will ignore this contribution.

It remains for us to determine the single string
partition function (dropping the $Thermal$ superscript), 
\begin{eqnarray}
  \label{eq:7}
  Z_1(a,b;\mu) = \tr_{\cH} e^{-bp_--ap_+}~,
\end{eqnarray}
for the IIB pp-wave string. 
This is straightforwardly achieved, as it is known how to quantize
the IIB pp-wave string in the lightcone gauge
\cite{Metsaev:2001bj}.
In this gauge, worldsheet diffeomorphism invariance is used to set
$X^+(\sig,\tau)=x^+ + (2\pi\alp\pr p^+)\tau$, and the theory is 
divided into sectors of fixed $p_- =
p^+$. After solving for the Virasoro constraints, the remaining transverse
degrees of freedom are eight {\em free} worldsheet
supersymmetric scalar multiplets with mass $\m=\mu\alp\pr p^+$ and
$p_+ =(2\pi)\inv \int d\sig P_+ $  as lightcone Hamiltonian;  
\begin{eqnarray}
  \label{eq:9}
  p_+ \equiv H_{lc}^{(\m)} 
 &=& \ove{\alp\pr p_-} \left(\ome_0 (N_0^B+ N_0^F)+ \sum_{n \geq 1}
  \ome_n^{(\m)} 
  (N_n^B+N_n^F+\tilde{N}_n^B+\tilde{N}_n^F)\right)
\end{eqnarray}
with
\begin{eqnarray}
\label{eq:9a}
\ome_n^{(\m)} =\mbox{sign}(n)\sqrt{n^2+\m^2}~.
\end{eqnarray}
$N^{B,F}_n$ and $\tilde{N}_n^{B,F}$ are the number operators for the eight
transverse right- and left-moving bosonic and fermionic oscillator
modes, and the right- and  leftmoving zero modes are identified as usual. 
As the (lightcone) 
theory is worldsheet supersymmetric the zero point energy
cancels between 
the bosons and fermions.

The choice of lightcone gauge solves the two worldsheet
reparametrization constraints explicitly. One  
single consistency condition
remains to be imposed. This is the descendant of the relation
\begin{eqnarray}
  \label{eq:14}
X^-(\sig,\tau)=X^-(\sig+2\pi,\tau) &\Leftrightarrow&  \alp^-_0 =
\tilde{\alp}_0^-
\end{eqnarray}
and becomes the level-matching constraint on the transverse
oscillators. The geometric reason behind this constraint is, as
always, 
the circle isometry of the worldsheet. The generator of
worldsheet translations in the sigma directions, normalized to
integer eigenvalues, is formally unchanged and
 equals $P=\sum_{n \geq 1} n (N_n-\tilde{N}_n)$. 
  Imposing
this constraint on the space of states by the introduction of a
delta function, the single string partition function equals
\begin{eqnarray}
  \label{eq:15}
  Z_1(a,b;\mu) = \int_0^\infty dp_- \int_{-\hlf}^\hlf d\tau_1\,
  e^{-bp_-}  
  z_{lc}(\tau_1, \frac{a}{2\pi\alp\pr p_-};\m=\mu\alp\pr p_-)~,
\end{eqnarray}
where $z_{lc}(\tau_1,\tau_2;\mu\alp\pr p_-)$ is 
the lightcone partition function 
\begin{eqnarray}
  \label{eq:35}
  z_{lc}(\tau_1,\tau_2;\m) \equiv \tr_{states}
  e^{-2\pi \tau_2(\alp\pr p_- H_{lc}^{(\m)})+2\pi i\tau_1 P} ~.
\end{eqnarray}
Note that we have chosen to define the lightcone partition
function such that the mass $\m=\mu\alp\pr p_-$ and the modular
parameter $\tau_2= a/2\pi\alp\pr p_-$ are independent.
The steps to evaluate eq. (\ref{eq:35}) are familiar. The bosonic and
fermionic sectors separate and each can be expressed in terms of the
building blocks \cite{Takayanagi:2002pi} (see also \cite{green})
\begin{eqnarray}
  \label{eq:36}
&& \hspace{-.5in} \The_{\alp,\del}(\tau,\bar{\tau};\m) = \non
&& \hspace{-.3in} e^{4\pi\tau_2
  E_c^\del(\m)}\prod_{n=-\infty}^\infty (1-
  e^{-2\pi\tau_2|\ome_{n+\del}|+2\pi i\tau_1(n+\del)+2\pi i \alp})(1-
  e^{-2\pi\tau_2|\ome_{n-\del}|+2\pi i\tau_1(n-\del)-2\pi i \alp})~,
\end{eqnarray}
where we have chosen as argument the standard
complex combination of the modular parameters $\tau=\tau_1+i\tau_2$.
Note that despite this notation 
the building blocks are not analytic in $\tau$.
The argument of the exponential prefactor, $E^{\del}_c$, is the zero
point energy shift or
Casimir energy for a chiral 
(one quarter of a complex two-dimensional) 
boson of mass $\m$ with periodicity $\phi(\sig+2\pi,t)= e^{2\pi i \del}
\phi(\sig,t)$,  
\begin{eqnarray}
  \label{eq:20}
  E^{\del}_c(\m) &=& 
  \frac{1}{4} \int_{-\infty}^{\infty} dk \,\,\ome(k) -\frac{1}{4}\sum_{n=-\infty}^\infty
|\ome_{n \pm \del}|  ~~~;~~~~~\ome(k)
  =\sqrt{k^2+\m^2} \non
 &=& -\ove{2\pi^2} \sum_{p=1}^{\infty} \int_0^{\infty} ds\,\,
  e^{-p^2s -\frac{(\pi\m)^2}{s}} \cos(2\pi \del p)~.
\end{eqnarray}
In terms of these building blocks the lightcone partition function
with the 
value of $\tau$ appropriate for eq. (\ref{eq:15}), 
$\tau =\tau_1+ia/2\pi \alp\pr p_-$, equals
\begin{eqnarray}
  \label{eq:22}
  z_{lc}(\tau,\bar{\tau};\mu\alp\pr p_-) = \int d\lam d\m \left[\frac
  {\The_{\hlf,0}(\tau,\bar{\tau};\m)}
  {\The_{0,0}(\tau,\bar{\tau};\m)}\right]^4  
e^{2\pi i\lam(\m-\mu\alp\pr p_-)}~.
\end{eqnarray}
The integration over the Lagrange multiplier $\lambda$ enforces the
proper value for the mass $\m$. 
Inserting this expression into eq. (\ref{eq:6}) and changing
integration variables to
$\tau_2 = ar/2\pi\alp\pr p_-$ we find the following  expression
for the free energy of an ideal gas of pp-wave strings 
\begin{eqnarray}
  \label{eq:8}
  -\bet F &=& \ln \cZ(a,b;\mu) \\ \nonumber
  &=& - \frac{a}{2\pi\alp\pr}
\int d\lam d\m \int_{-\hlf}^{\hlf} d\tau_1
   \int_0^{\infty} d\tau_2 \ove{\tau_2^2}\left[\frac
  {\The_{\hlf,0}(\tau,\bar{\tau};\m)}
  {\The_{0,0}(\tau,\bar{\tau};\m)} \right]^4  
   \sum_{r={\rm odd}}^{\infty} 
   e^{-\frac{abr^2}{2\pi\alp\pr \tau_2} +2\pi i\lam (\m - \mu ar/2\pi \tau_2)}
~.
\end{eqnarray}

\section{High temperature behaviour of pp-wave strings}
\setcounter{equation}{0}

Having computed the partition function of the IIB pp-wave string, we
may start to analyze its thermodynamic properties.
Particular questions of importance are:
\begin{itemize}
\item Does the IIB pp-wave string have a Hagedorn temperature (do
  level densities grow exponentially)?
\item If so, does the Hagedorn behaviour of the IIB signal the onset
  of a 
  phase transition, or is there a limiting temperature?
\end{itemize}
As the difference between the IIB pp-wave string 
and the Minkowski string is
rather mild, we expect that the gross characteristics are
unchanged. Naively both questions should therefore receive a positive
answer, and the spectrum of the IIB pp-wave string appears to support
this conclusion.  
For highly excited modes with individual
frequency $\ome_n = \sqrt{n^2 +(\mu\alp\pr p_-)^2} = n(1+
\frac{(\mu\alp\pr p_-)^2}{2n^2} + \ldots)$ a given (quantized)
energy level $E=N$ 
is somewhat less easily partitioned, but asymptotically (where $\mu$
can effectively be ignored) it certainly approaches the exponentially
growing Minkowski spectrum. The pp-wave density
of states is therefore also expected to 
grow exponentionally, albeit less rapidly at high $T$ than the flat space
string. There should therefore exist 
a pp-wave Hagedorn temperature, whose
value is somewhat higher than that of the Minkowski space
string. 

By the same reasoning, one is inclined to associate
the Hagedorn temperature with a phase transition
and not with a limiting temperature. The parameter which
determines the nature of the Hagedorn behaviour, limiting or phase
transition, is the subleading
power $\gam$ in the asymptotic density of states \cite{dienes}
\begin{eqnarray}
  \label{eq:37}
d(E \rar \infty)  ~\sim~  E^{\gam} e^{E/T_H}~.
\end{eqnarray}
Again, one expects that asymptotically the effect of the 
pp-wave deformation,
characterized by the value of $\mu$, on the coefficient $\gam$ 
is negligible. Surprisingly we will find that this is not the case.
A closer look reveals why.  The
effective mass in the lightcone gauge is $\m=\mu\alp\pr p_-$ rather
than $\mu$. Although asymptotic level densities $d(n)$ are clearly
unaffected by $\mu$, the density of states $d(E)$ at high
energies (equivalent to large lightcone momentum $p_-$) may not be
insensitive to the deformation. We will show that the effect of $\mu$
remains mild in that the growth of states is still exponential ---
there is a Hagedorn temperature --- and just mild
enough so that the pp-wave string also
undergoes a phase 
transition at $T=T_H$, as recently shown by 
\cite{blt}.\footnote{In a preprint version, our calculations suggested that the
Hagedorn temperature was limiting, a result we conjectured
to be an artifact of approximations being made. Indeed, as
the recent new results of \cite{blt} show, this is the case; 
see endnote for details.}

\subsection{Hagedorn behaviour}

Let us first show that there is a Hagedorn temperature at which the
free energy diverges. This happens when the integrand of
eq. (\ref{eq:8}) diverges exponentially for $\tau_2 \rar 0$. 
Hence we need to compute the
asymptotic behaviour of the building blocks 
$\The_{\alp,\del}(\tau,\bar{\tau};\m)$ as $\tau \rar 0$. As we already
 briefly noted, this has to
be done with some care as the effective mass $\m$ is constrained to 
equal $\m = \mu ar/2\pi\tau_2$; see eq. (\ref{eq:8}). Even so, the effect of
$\mu$ on the 
asymptotic level densities of the string, we have argued, should be
mild, and we therefore make the
ansatz that for $\tau_2 \rar 0$, with $\m\tau_2 = \tilde{\mu}$ fixed,
the building blocks 
$\The_{\alp,\del}(\tau,\bar{\tau};\m)$ diverge exponentially; 
\begin{eqnarray}
  \label{eq:10}
  \The_{\alp,\del}(\tau,\bar{\tau};\frac{\tilde{\mu}}{\tau_2})~ \sim~
  \exp \left[\frac{\xi(\tilde{\mu})}{\tau_2}\right] 
  ~~~\mbox{as}~~\tau_2 \rar 0~.
\end{eqnarray}
To compute the coefficient $\xi(\tilde{\mu})$, consider the logarithm
of 
eq. (\ref{eq:36})
\begin{eqnarray}
  \label{eq:17}
  \ln \The_{\alp,\del}(\tau,\bar{\tau},\frac{\tilde{\mu}}{\tau_2}) &=&
 4\pi \tau_2 E_c^{\del}(\frac{\tilde{\mu}}{\tau_2}) +  
 \sum_{n =-\infty}^{\infty} 
\ln(1- e^{-2\pi\tau_2|\ome_{n+\del}|+2\pi i \tau_1 (n+\del)+2\pi i
 \alp 
}) +{\rm c.c.}
\end{eqnarray}
The coefficient $2\pi\tau_2 |\ome_{n+\del}|$ equals 
\begin{eqnarray}
\label{eq:17a}
2\pi\tau_2
|\ome_{n+\del}| =
2\pi\tilde{\mu}\sqrt{1+\frac{(n+\del)^2\tau_2^2}{\tilde{\mu}^2}}~.   
\end{eqnarray}
In the limit $\tau_2 \rar 0$ we may replace the sum over $n$ by an
integral, provided we keep the ratio $\tau_1/\tau_2 \equiv \th$ fixed
(The limit is therefore equal to $|\tau| \rar 0$, $\arg(\tau)$
fixed. If we keep $\tau_1$ finite instead, the limit $\tau_2 \rar 0$
is convergent).
\begin{eqnarray}
  \label{eq:23}
&& \sum_{n=-\infty}^{\infty}\ln(1-
e^{-2\pi\tilde{\mu}\sqrt{1+{(n+\del)^2\tau_2^2}/{\tilde{\mu}^2}}+2\pi i
\th\tau_2 (n+\del)+2\pi i \alp
}) ~~~\stackrel{\tau_2 \rar 0}{\longrightarrow} \non
&& \hspace{1in}\frac{\tilde{\mu}}{\tau_2}
\int_{-\infty}^{\infty} {dn} \ln(1-
e^{-2\pi\tilde{\mu}\sqrt{1+{n^2}}+2\pi i \tilde{\mu}\th n  +2\pi i \alp
}) ~~\equiv~~  - \frac{\tilde{\mu}}{|\tau|} f(\tilde{\mu},\th,\alp)~,
\end{eqnarray}
and we have determined the coefficient of divergence 
$\xi(\tilde{\mu})$ implicitly in terms of an unknown function $f(\tilde{\mu},\th,\alp)$
\begin{eqnarray}
  \label{eq:28}
  \xi(\tilde{\mu}) = - \frac{\tilde{\mu}\tau_2}{|\tau|}\left[f(\tilde{\mu},\th,\alp)+\bar{f}(\tilde{\mu},\th,\alp)\right]~.
\end{eqnarray}
That the divergence of the building blocks
$\The_{\alp,\del}(\tau,\bar{\tau};\m)$ depends on the direction with
which one approaches $\tau=0$, is reflective of 
their non-analyticity.

This function $f(\tilde{\mu},\th,\alp)$ is formally equal to
\begin{eqnarray}
\label{eq:23a}
f(\tilde{\mu},\th,\alp) 
&=& 
\frac{|\tau|}{\tau_2}\sum_{\ell=1}^{\infty} \int_{-\infty}^{\infty} dn\,
\frac{e^{-2\pi \ell
  \tilde{\mu}\sqrt{1+n^2}+2\pi i \ell \tilde{\mu}\th n +2\pi i \alp\ell}}{\ell}
\non
&=& 2 \sum_{\ell=1}^{\infty} \frac{e^{2\pi i \ell \alp}}{\ell}
K_1(2\pi \ell \tilde{\mu} \sqrt{(1+\th^2)})   ~,
\end{eqnarray}
where $K_1(x)$ is the modified Bessel function of the second kind
\cite{abram}. 
For $\alp=0$, $f(\tilde{\mu},\th,0)= f(\tilde{\mu}\sqrt{1+\th^2},0)$
is a monotonic function which for small and large values of its
argument asymptotes to
\begin{eqnarray}
  \label{eq:24}
f(x,0) &\equiv& 2 \sum_{\ell=1}^{\infty} \frac{K_1(2\pi \ell x)}{\ell}
  ~, \non
  x \rar 0 ~&:&~~~  f(x,0) \sim \frac{\pi}{6x}~, \non
  x \rar \infty ~&:&~~~f(x,0) \sim  \frac{e^{-2\pi x}}{\sqrt{x}}~.
\end{eqnarray}
For $\alp=\hlf$, $f(x,\hlf)$
behaves as
\begin{eqnarray}
  \label{eq:16}
  f(x,\hlf) &\equiv& 2 \sum_{\ell=1}^{\infty} (-1)^{\ell}
\frac{K_1(2\pi \ell x)}{\ell} ~,\non
  x \rar 0 ~&:&~~~  f(x,\hlf) \sim - \frac{\pi}{12 x} ~,\non
  x \rar \infty ~&:&~~~f(x,\hlf) \sim - \frac{e^{-2\pi x}}{\sqrt{x}}~.
\end{eqnarray}
Note that
$\bar{f}(x,\alp)=f(x,-\alp)
=f(x,-\alp+1)$.

Collecting this information, we establish the Hagedorn temperature as
the combination of parameters $a,b$ which fail to dampen the
exponential growth of oscillator states 
in the UV. Integrating out the Lagrange multiplier $\lam$ in
eq. (\ref{eq:8}) and substituting the proper value for the mass $\m$, 
the exponential growth equals
\begin{eqnarray}
  \label{eq:40}
\hspace{-.2in}  \left[\frac
  {\The_{\hlf,0}(\tau,\bar{\tau};\frac{\mu a r}{2\pi \tau_2})}
  {\The_{0,0}(\tau,\bar{\tau};\frac{\mu a r}{2\pi\tau_2} )} \right]^4
  \stackrel{\tau_2 \rar 
  0}{\longrightarrow}~ \exp\left[ \frac{8 \mu a r
  }{2\pi|\tau|} \left(f(\frac{\mu a r}{2\pi}\sqrt{1+\th^2},0)
 - f(\frac{\mu a r}{2\pi}\sqrt{1+\th^2},\hlf)\right) \right]~.
\end{eqnarray}
Because $f(x,\alp)$
tends very rapidly to zero for increasing $x$, 
the leading divergent term in the
integrand of eq. (\ref{eq:8}) is due to the single string contribution
for which $r=1$. Thus for small values of $\tau_2$
the integrand is dominated by an exponential factor 
\begin{eqnarray}
  \label{eq:25}
   \exp(-\frac{ab}{2\pi\alp\pr\tau_2}) \exp\left[ \frac{8 \mu a 
  }{2\pi|\tau|} \left(f(\frac{\mu a }{2\pi}\sqrt{1+\th^2},0)
 - f(\frac{\mu a }{2\pi}\sqrt{1+\th^2},\hlf)\right) \right]~,
\end{eqnarray}
and the partition function will be convergent if the parameters $a, b$
which determine the temperature are chosen such that the argument in
(\ref{eq:25}) is negative. This is only the case when
\begin{eqnarray}
  \label{eq:26}
  b > 8 \frac{\tau_2\mu\alp\pr}{|\tau|}\left( f(\frac{\mu
  a}{2\pi}\sqrt{1+\th^2},0)-  f(\frac{\mu
  a}{2\pi}\sqrt{1+\th^2},\hlf))\right) ~. 
\end{eqnarray}
The explicit $\th \sim \tau_1$ dependence may appear curious, but, as
we mentioned previously, this is a consequence of the non-analytic
nature of the building blocks. In the final expression for the free
energy $\tau_1$ is integrated over and no dependence remains. To
explicitly perform the $\tau_1$ integral is not
straightforward. Inspection of the RHS of eq. (\ref{eq:26}), however,
shows that it is a monotonically decreasing function of $\theta$. 
This allows us to
argue that a minimum requirement for convergence is eq. (\ref{eq:26})
with $\th$ set to 
zero,\footnote{For an important subtlety regarding this reasoning, see note
added in proof.}
\begin{eqnarray}
  \label{eq:45}
  b > 8 \mu\alp\pr \left( f(\frac{\mu a}{2\pi},0)-
  f(\frac{\mu a}{2\pi},\hlf))\right) ~. 
\end{eqnarray}
As expected the pp-wave string therefore has a Hagedorn temperature,
which occurs when this inequality is saturated. The physical
temperature measured by different observers (parameterized by $a$) 
at the bound equals
\begin{eqnarray}
  \label{eq:44}
  T_{H}(a;z)^{-2} = 8
  a\mu\alp\pr
  \left(f(\frac{a\mu}{2\pi},0)-f(\frac{a\mu}{2\pi},\hlf)\right)+ 
  \mu^2z^2 a^2~.
\end{eqnarray}
For $\mu=0$, this indeed gives 
the standard, observer independent, IIB Hagedorn temperature
\begin{eqnarray}
  \label{eq:27}
  ab = T_H^{-2} = 8\pi \alp\pr (\frac{\pi}{6} +\frac{\pi}{12}) = 4\pi^2
  \alp \pr~,
\end{eqnarray}
whereas for $\mu \neq 0$ the rapid vanishing of $f(x)$ implies that
the Hagedorn temperature rises quickly with $\mu$, but only disappears
for $\mu$ strictly infinite. 

Note finally that in
  all the expressions above 
only those combinations of $a,~b$ and $\mu$
  appear which are invariant under the Lorentz boost plus rescaling:
  $a 
  \rar \alp a$,~$b \rar b/\alp$ and $\mu \rar \mu/\alp$, as we argued
  in section 2.

\subsection{Phase transition or limiting temperature}

The crucial question is whether the Hagedorn temperature
$T_H$ is limiting or whether the divergence of the partition function
for $ T> T_H$
signals the onset of a phase transition. The distinction between the
two modes of behaviour is whether thermodynamic quantities at $T=T_H$
are finite or not. We will focus on the free energy, $\bet F = -\ln {\cal Z}$, 
eq. (\ref{eq:8}). A necessary, but not sufficient, requirement for the
possibility of a phase transition is that the free energy is finite at
$T=T_H$. To establish whether this is the case, 
we also need to know the subleading divergence
of the building blocks $\The_{\alp,\del}(\tau,\bar{\tau};\m)$ as
$\tau_2 \rar 0$. Postulating 
that the behaviour of $\The_{\alp,\del}(\tau,\bar{\tau};\m)$ is
of the form
\begin{eqnarray}
  \label{eq:49}
  \The_{\alp,\del}(\tau,\bar{\tau};\frac{\tilde{\mu}}{\tau_2}) \sim
  \tau_2^{c(\tilde{\mu})} e^{\frac{\xi(\tilde{\mu})}{\tau_2}}~, 
\end{eqnarray}
the free energy, eq. (\ref{eq:8}), will be finite at $T=T_H$ if
$c(\mu,\alp=\hlf)-c(\mu,\alp=0)$ is large enough to overcome the
  $\tau_2^{-2}$ factor in 
the measure.\footnote{Applying these observations to determine the nature of the Hagedorn temperature
depends sensitively on working with the full integral instead of just the integrand of the
partition function; see endnote. As \cite{blt} shows, 
the $\tau_1$ Lagrange multiplier 
integral contributes an additional factor
$\tau_2^{3/2}$ to the measure. Hence if $c(\mu,\alp=\hlf)-c(\mu,\alp=0)
> -\ove{8}$, the free energy is finite and  
the Hagedorn bound demarcates a
phase transition. \label{note:1}} 
 This is the case if
  $c(\mu,\alp=\hlf)-c(\mu,\alp=0) > \ove{2}$. The naive
expectation --- which we will show {\em not} to be true --- 
is that $c(\mu)$ is a
smooth function with $c(\mu=0;\alp\neq 0)=0;~c(\mu=0;\alp=0)= -1/2$.

We can compute the power $c$ by taking the logarithm as before,
\begin{eqnarray}
  \label{eq:50}
  \ln \The_{\alp,\del}(\tau,\bar{\tau};\frac{\tilde{\mu}}{\tau_2})  -
  \frac{\xi(\tilde{\mu})}{\tau_2} = c(\tilde{\mu})\ln \tau_2 +
  \cO(\tau_2)~. 
\end{eqnarray}
Recalling the determination of $\xi(\tilde{\mu})$ 
from eq. (\ref{eq:23}), the LHS
of this equation equals
\begin{eqnarray}
  \label{eq:51}
\nonumber
  \sum_{n=-\infty}^{\infty}\ln(1-
  e^{-2\pi\tilde{\mu}\sqrt{1+{(n+\del)^2\tau_2^2}/{\tilde{\mu}^2}}
  +2\pi  
  i \th\tau_2 (n+\del)+2\pi i \alp 
}) - \frac{\tilde{\mu}}{\tau_2}
\int_{-\infty}^{\infty} {dn} \ln(1-
e^{-2\pi\tilde{\mu}\sqrt{1+{n^2}}+2\pi i \tilde{\mu}\th n  +2\pi i
  \alp 
})~.
\end{eqnarray}
Setting $\th=0$ (basing ourselves on the arguments below
eq. (\ref{eq:26})\footnote{This is allowed, if we correctly account
  for any contributions to the $\tau_2$ 
measure, see footnote \ref{note:1}.})
and using the Poisson resummation identity
\begin{eqnarray}
  \label{eq:52}
  \sum_{n=-\infty}^{\infty} f(n) = \sum_{k=-\infty}^{\infty}
  \int_{-\infty}^{\infty} dx \, e^{2\pi i k x} f(x)~,
\end{eqnarray}
this can be simplified to 
\begin{eqnarray}
  \label{eq:53}
  2 \sum_{k=1}^{\infty} \int_{-\infty}^{\infty} dn \ln (1-e^{-2\pi
  \tilde{\mu} \sqrt{ 1+ n^2\tau_2^2/\tilde{\mu}^2} +2\pi i \alp})
  e^{2\pi i k (n-\del)} {=} c\ln\tau_2 + \cO(\tau_2)~.
\end{eqnarray}
Formally the LHS equals
\begin{eqnarray}
  \label{eq:55}
 - 2 \sum_{k=1}^{\infty} \sum_{\ell =1}^{\infty} \frac{e^{2\pi i (\ell
   \alp-k\del)}}{\ell} \frac{\tilde{\mu}}{\tau_2} \frac{K_1(2\pi \ell
   \tilde{\mu}
   \sqrt{1+({k}/{\ell\tau_2})^2})}{\sqrt{1+(k/\ell\tau_2)^2}}~.  
\end{eqnarray}

Evaluating eq. (\ref{eq:55}) in the limit $\tau_2 \rar 0$, we
may approximate the Bessel function
$K_1(2\pi\ell\tilde{\mu}\sqrt{1+(k/\ell\tau_2)^2})$ with its 
asymptote at infinity,
\begin{eqnarray}
  \label{eq:21}
  \tau_2 \rar 0:~~~~~{\rm LHS_{\mbox{\small (\ref{eq:53})}}} ~=~ 
-2 \sum_{k,\ell=1}^{\infty}
\frac{\tilde{\mu}e^{2\pi i (\ell \alp-k\del)}}{ \sqrt{\ell^2\tau_2^2+k^2}}
  \frac{e^{ - 2\pi\ell\tilde{\mu}\sqrt{1+(k/\ell\tau_2)^2}}}{\sqrt
  { 4\ell\tilde{\mu}\sqrt{1+(k/\ell\tau_2)^2}}} ~\rar~ 0~,
\end{eqnarray}
which goes to zero exponentially fast. We find {\em no} logarithmic
divergence and conclude that $c(\tilde{\mu})=0$ for all values of
$\alpha$ and $\del$. This establishes our
claim. For the implication is that the free energy is {infinite}
at $T=T_H$ and there is {no} phase transition, contrary to our
intuition based on the flat space, $\mu=0$ results.\footnote{See
  footnote \ref{note:1}. With the improved treatment of the $\tau_1$
  integration of \cite{blt}, one concludes that the Hagedorn bound
  demarcates a phase transition.}

This answer is surprising, however. How does it  
mesh with the known result for $\mu=0$? Inspection shows
that in eq. (\ref{eq:55}) the limits $\mu \sim  \tilde{\mu} \rar 0$
and $\tau_2 \rar 0$ do not commute. 
If we take the limit $\tilde{\mu} \rar 0$
first, we may approximate the Bessel function
$K_1(2\pi\ell\tilde{\mu}\sqrt{1+(k/\ell\tau_2)^2})$  
with its asymptote
near zero, instead of the asymptote near infinity 
\begin{eqnarray}
  \label{eq:30}
  \tilde{\mu} \rar 0:~~~~~{\rm LHS_{\mbox{\small (\ref{eq:53})}}} ~=~-2\sum_{k,\ell=1}^{\infty}
  \frac{\tilde{\mu}e ^{2\pi i (\ell \alp-k\del)}}{\sqrt{\ell^2\tau_2^2+k^2}}
  \frac{1}{2\pi\ell \tilde{\mu}\sqrt{1+(k/\ell\tau_2)^2}}~.
\end{eqnarray}
The dependence on $\tilde{\mu}$ drops out, and one is tempted to
conclude that
$c(\tilde{\mu}=0)$ also vanishes. However, iff $\alpha =0$ and
$\del=0$, the 
resulting double
sum is divergent. One is not allowed
to take the limit $\tilde{\mu} \rar 0$ within the sum for
$\alp,\del=0$, and it 
follows that
our unstated assumption that
$c(\tilde{\mu}; \alpha,\del=0)$ is a smooth function of $\tilde{\mu}$
is incorrect.
A careful analysis, which we have included in an appendix for completeness, yields the
known value $c(\tilde{\mu}=0; \alpha,\del=0)=-\hlf$.

This incompatibility of limits is 
already evident in the expression
for the frequency $\omega_n =\sqrt{n^2+(\mu\alpha p_{-})^2}$. For $\mu
\neq 0$ and only for $\mu \neq 0$, in the high energy limit $p_{-} \rightarrow \infty$, 
all oscillator levels can be populated at no cost in extra energy.
Pursuing the argument put forth in the beginning of this section to
the end, one would conclude that this costless 
contribution from the tower of string states 
gives rise to a divergent free energy.
Hence, the Hagedorn temperature is a limiting
temperature.\footnote{This intuitive picture is naive. As we
  conjecture later on, and \cite{blt} has recently been able to show,
  the Hagedorn bound ought to, and does, signal a phase transition.}

We thus see that the Hagedorn behaviour of the $\mu=0$ IIB 
Minkowski
  string 
  is fundamentally different from that of the IIB pp-wave string; 
there is no
  continuum limit between the two. There is in fact a good 
physical explanation
  for why our intuition that at high energies the $\mu$ dependence of the
background
  is irrelevant, and that the presence of $\mu$ should only affect the
  IR behaviour, fails. 
String theory is modular
invariant, which means that UV physics ``knows'' 
about IR physics. This also holds for the pp-wave string. Indeed,
``strikingly'', despite their nonanalyticity, the building blocks
$\The_{\alp,\del}(\tau,\bar{\tau};\frac{\tilde{\mu}}{|\tau|})$
have nice modular properties \cite{Takayanagi:2002pi,green}. Under the
$S$-modular 
transformation, which mixes UV and IR physics, they transform as
\begin{eqnarray}
  \label{eq:31}
  \The_{\alp,\del}(\tau,\bar{\tau};\frac{\tilde{\mu}}{|\tau|}) =
\The_{-\del,\alp}(-\ove{\tau},-\ove{\bar{\tau}},\tilde{\mu})~.
\end{eqnarray}
The exact asymptotics of $\The_{-\del,\alp}(\tau,\bar{\tau};\tilde{\mu})$
at $\tau_2 \rar \infty$, including any power law divergence, are easily read off
from the defining
expression (\ref{eq:36}) (based on our previous arguments, $\tau_1$ is
again set to zero from here on)
\begin{eqnarray}
  \label{eq:32}
  \lim_{\tau_2 \rar \infty}
  \The_{-\del,\alp}(\tau,\bar{\tau};\tilde{\mu}) = \exp(4\pi \tau_2
E_c^{\alp}(\tilde{\mu}))~.
\end{eqnarray}
Hence the exact $\tau_2 \rar 0$ divergence of
$\The_{\alp,\del}(\tau,\bar{\tau};\frac{\tilde{\mu}}{\tau_2})$ equals
\begin{eqnarray}
  \label{eq:38}
  \lim_{\tau_2 \rar
  0}\The_{\alp,\del}(\tau,\bar{\tau};\frac{\tilde{\mu}}{\tau_2})
  = \exp(4\pi \frac{E^{\alp}_c(\tilde{\mu})}{\tau_2})~.
\end{eqnarray}
This is indeed
equal to the expression we had before (eq. (\ref{eq:28})), and
illustrates how the high 
energy physics knows about the
infrared: the coefficient of exponential growth is the 2D 
Casimir energy.
This 
unambiguously determines that the power law divergence coefficient 
$c(\mu \neq 0) = 0$.

The essence of why this does not match with 
the flat space result $c(\mu=0;\alp=0) = -\hlf$, 
resides in the modular properties of
$\The_{0,0}(\tau,\bar{\tau};\m)$. This building block strictly
vanishes in the limit $\mu \rar 0$. As shown in 
\cite{green}, 
the modular properties of the corresponding flat space building 
block, the Dedekind eta function, which include a power law factor, 
are obtained at the subleading level in $\mu=0$ from 
the modular properties of the building block
$\The_{0,0}(\tau,\bar{\tau};\m)$. The vanishing of
$\The_{0,0}(\tau,\bar{\tau};\m)$ for $\mu \rar 0$ is due to the
appearance of zero-modes. Naively at
high energies the presence or absence of 
zero modes, an infrared issue, should not matter. What the computation
explicitly shows, is that the effective mass in the problem is 
$\m= \tilde{\mu}/\tau_2$, rather than $\mu$.
Thus in the naive UV, $\tau_2 \rar 0$, 
the mass blows up, and cannot be
ignored.\footnote{Oddly enough,
  it  is the
$S$-transformed part of
  the building block $\tau_2 \rar 
\infty$, the naive IR region, where the effect of the 
mass $\mu$ is the mildest.}  
This is the reason why the Hagedorn behaviour of the pp-wave
string is 
different from that of the flat space string.\footnote{For clarity, the issue we are
referring to here is how in string theory IR zero modes impact the UV 
Hagedorn behaviour and the different zero mode structures 
of the pp-wave
string compared to the Minkowski string theory. 
 The pp-wave string behaves as a $T^8$
compactification. Whether the Hagedorn temperature is
limiting or signals a phase transition, depends on one further step of
analysis.} 

\section{Implications for the CFT dual}
\setcounter{equation}{0}

\label{sec:impl-cft-dual}

Pp-wave strings can be derived as a scaling limit from AdS strings,
and through this limit they inherit a duality map with a subsector of
a CFT \cite{Berenstein:2002jq}. They are the only known example of a
gauge-theory/string theory duality where the string side is
understood, and as such hold promise in both explaining what stringy
effects are in gauge theory, as well as using gauge theory to answer 
string theory questions. For the IIB pp-wave string the precise map is
to a double scaling limit of $\cN=4$, $SU(N)$ Super-Yang-Mills (SYM),
the BMN limit, which sends both the rank of the gauge group $N$ and the
$R$-charge
$J$ to infinity, keeping the ratio $J^2/N$ fixed
\cite{Berenstein:2002jq,Gubser:2002tv,Kristjansen:2002bb,Berenstein:2002sa,Gross:2002su,Constable:2002hw,Gopakumar:2002dq,Verlinde:2002ig}.

The symmetry charges of both theories 
are mapped to each other as follows 
\begin{eqnarray}
  \label{eq:41}
  \frac{2p_+}{\mu} = E - J~~,~~~~2\mu\alp\pr p_- =
  \frac{J}{\sqrt{\lam}}~,~~~~~\lam = g^2_{YM}N=4\pi g_sN~.
\end{eqnarray}
Here $E$ is the energy of the theory on $S^3$ (or conformal weight on $R^4$),
and $J$ a generator of a
$U(1)_R$ subgroup of $SU(4)_R$. The string partition function,
$\cZ(a,b;\mu)$ we computed in eq. (\ref{eq:8}), is a function of the
two independent variables, $a,~b$, thermodynamically conjugate to the operators
$p_+,~p_-$ (recall that $\mu$ can be removed by the Lorentz boost). 
One linear combination of these is the temperature of the
string heat bath,
the other designates the family of observers which perform the
measurement. From the SYM point of view we see that we
are computing the partition function at 
a finite temperature $T_{YM}$ and finite $R$
charge chemical potential $\nu_R$. To determine the precise relation
between $a,~b$ and $T_{YM}$ and $\nu_R$, we need to recall some facts
about both
theories and the map in more detail. 

The derivation of the duality between pp-wave strings
and the BMN doubly scaling limit of $\cN=4$ SYM starts with a global
coordinate choice for the AdS factor. This means we are
considering the SYM theory on 
$S_3$ rather than $\RR^4$. The finite temperature, finite
density SYM partition function is therefore well 
defined, with the scale set by the radius of the $S_3$. A second
detail concerns thermodynamics in a system with non-zero groundstate
energy. When we consider a system (ideal gas) at finite temperature, what we
really mean is that the constituents of the system 
 have an average {\em excitation} energy $E_{ex} \sim T$ above the
 groundstate. In other words the temperature is the thermodynamic
 conjugate to the excitation energy $E_{ex}=E-E_0$ rather than the
 absolute energy $E$ (see e.g. \cite{landau}). In $\cN=4$ SYM a state
 with a specific $R$-charge $J$ has at least an absolute energy $J$
 due to the BPS condition. Hence within a sector of fixed $J$, there
 is a groundstate energy, and when considering the thermodynamics
 within such a sector, the temperature is the conjugate variable to
 $E-J$. The corresponding statement on the pp-wave string side is of
 course that the light-cone Hamiltonian $p_+$ is positive semi-definite.

With this in mind, and the map of operators eq. (\ref{eq:41}), 
the identification of the string thermodynamic
variables in terms of those of $\cN=4$ SYM is straightforward.
The temperature and chemical potential equal 
\begin{eqnarray}
  \label{eq:39}
  T_{YM}\inv = \frac{a\mu}{2}~,~~~~~ \sqrt{\lam}\nu_R =
  \frac{b}{2\mu\alp\pr}~. 
\end{eqnarray}
However, we are not considering the full $\cN=4$ $SU(N)$ SYM theory
but only the 
physical degrees of freedom that survive 
the double scaling limit $N,J
\rar \infty$, with $J^2/N$ fixed. It is therefore more appropriate to
define a chemical potential $\tilde{\nu}\equiv \nu_R\sqrt{\lam}$ conjugate to
$\tilde{J}
\equiv 
J/\sqrt{\lam}$. Finally, all computations in the preceding
sections are done in the ideal gas approximation where $g_s \ll
1$. We should therefore keep in mind that the results obtained should
only apply when the Yang-Mills coupling is also vanishingly weak.

\subsection{Hagedorn bound for $\cN=4$ SYM}

We have seen that for an ideal gas of IIB 
pp-wave strings the free energy diverges at 
\begin{eqnarray}
  \label{eq:42}
  \frac{b}{8 \mu\alp\pr} =
  f(\frac{a\mu}{2\pi},0)-f(\frac{a\mu}{2\pi},\hlf) ~.
\end{eqnarray}
When the left-hand side is less than the right-hand side, the
(perturbative) IIB pp-wave string 
theory is does not exist, as the Hagedorn behaviour is 
limiting.\footnote{This is the conclusion we drew in a preprint version.  However, as shown 
in footnote \ref{note:1} and the note added in proof, using the 
improved calculation of \cite{blt} 
 the Hagedorn temperature is seen to demarcate a phase transition both
 for compactifications on $T^p$, $p<9$ and pp-wave strings, updating
 our and previous results.\label{note:3}}

In terms of SYM variables the 
Hagedorn bound reads
\begin{eqnarray}
  \label{eq:43}
 \tilde{\nu} = 4\left(f(\ove{\pi T},0)-f(\ove{\pi
 T}, \hlf)\right)~.
\end{eqnarray}
In figure 1, we have plotted the bound in the temperature-chemical potential
phase diagram of $\cN=4$ SYM.\footnote{Timelikeness of the Killing
  vector also constrains the variables $a$ and $b$. For a global
  Killing vector $a,~b >0$ simply implies that the SYM 
temperature and density are positive semi-definite.}
The existence of a limiting Hagedorn temperature in the dual pp-wave string 
would appear to say that the area of the phase diagram where the BMN
density $\tilde{\nu}$ is less than the RHS of eq. (\ref{eq:43}) is
inaccessible. This is peculiar, as it includes the 
familiar $\tilde{\nu} \sim \nu =0$ region of the theory. In fact
through 
the AdS/CFT correspondence quite
a lot has become known about the finite density, finite temperature
behaviour of large $N$, 
$\cN=4$ SYM through the study of spinning branes in
supergravity
\cite
{gubser,Cai:1998ji,Russo:1998by,Chamblin:1999tk,Harmark:1999xt,Evans:2001ab}.
Unfortunately these studies are not able to clarify what the physics is near
the
Hagedorn bound. They are performed in supergravity, the $\alp\pr \rar 0$ limit
of the AdS string theory, and form a complementary regime to that of the
pp-wave
\cite{Berenstein:2002sa}. 

\begin{figure}[htbp]
\centering
\begin{picture}(500,150)(-250,-50)
\put(-240,-40)
{ \includegraphics[scale=0.75]{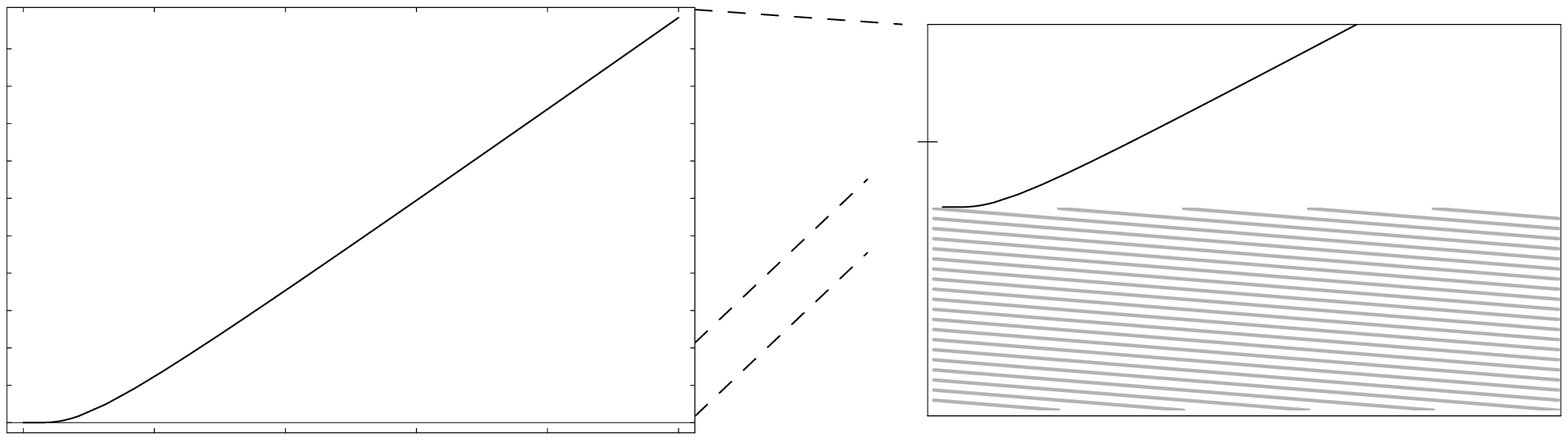}}     
\put(-105,15){{\underline{\em (Ideal Gas)}}}
\put(-130,0){{\underline{\em Inaccesible Area}}}
\put(-200,50){pp-wave String}
\put(-250,-5){$\uparrow$}
\put(-250,-20){$\tilde{\nu}_R$}
\put(0,-25){$\nu_R$}
\put(0,-15){$\uparrow$}
\put(-220,-50){$T_{YM}~\rar$}
\put(30,-50){$T_{YM}~\rar$}
\put(5,30){{$\scriptstyle \frac{1}{\sqrt{N}}$}}
\put(5,50){{$\scriptstyle \ove{\sqrt{\lam}}$}}
\put(-35,-08){{$\scriptstyle 1$}}
\put(-35,-25){{$\scriptstyle g_{YM}$}}
\put(-232,-25){{$\scriptstyle 0$}}
\end{picture}
  \caption{{\em Temperature-chemical potential 
    plot for $g^2_{YM} \ll 1$ 
    of the BMN limit of $\cN=4$ SYM on
    $S_3$. The right-hand-side shows where the $g^2_{YM} \ll 1$ 
    BMN limit region lies  within the
    full temperature-density plot of $\cN=4$ SYM. The shaded region,
    where $\nu$ is not parametrically comparable to $1/\sqrt{N}$ is
    excluded. This includes the giant graviton 
    phase where $\nu \sim \sqrt{\lam}/N$ \cite{Berenstein:2002sa}.}
}
  \label{fig:1}
\end{figure}

On the other hand the region of applicability of these studies is 
beyond the Hagedorn
bound and therefore, surely, this region is physically
accessible.  A possible
explanation for the apparent mismatch between the IIB pp-wave
string theory and $\cN=4$ SYM is that extending the usual zero
temperature  
duality may
be more subtle than the approach used in this
section.\footnote{Throughout our analysis we have worked in the
  context of the 
canonical ensemble; it would be worth checking that temperature
fluctuations are under control to ensure the validity of this
assumption. The recent paper \cite{blt} discusses this point.} 
Perhaps,
for example,  it
is only within the ideal gas approximation of vanishingly weak
coupling that the Hagedorn bound is limiting. Indeed with small, but
non-zero coupling, an infinite free energy in a gravitational
system is not likely to be a stable situation. At finite $g_s \lesssim
1$ the system will fall out of equilibrium long before one approaches
the Hagedorn bound. One suspects that at finite string coupling,
multistring bound states 
and/or nonperturbative states, such as D-brane black holes,
become dominant at high energies, whereas in the
ideal gas approximation the high energy behavior is effectively that
of a single string. Perhaps, then,  the Hagedorn bound 
reflects a breakdown of the 
duality between {\em ultra perturbative} (perturbative string theory in the ideal gas
approximation) 
IIB pp-wave string theory and $\cN=4$ SYM.\footnote{This conjecture of
  ours, that the Hagedorn
bound should demarcate a phase boundary, has indeed been 
shown correct by Brower, Lowe
and Tan \cite{blt}. They are able to 
come to this conclusion by including contributions to the integrand of
the partition function arising from the Lagrange multiplier $\tau_1$
integral; see note added in proof. \label{note:2}}

\section{Conclusion}

We have studied the finite temperature behavior of IIB strings in a pp-wave 
background and found, as expected from flat-space experience, a Hagedorn 
temperature. However, whereas the Hagedorn temperature in Minkowski space  
indicates the onset of a phase transition, our calculations show the ideal gas 
pp-wave Hagedorn temperature to have a different character: it is a limiting 
temperature.\footnote{See footnotes \ref{note:3} and \ref{note:2}.} 
At first this might seem surprising since one would
  have thought that
the pp-wave spacetime deformation becomes irrelevant at high
energies; the high temperature pp-wave
string should be smoothly 
connected to the high temperature Minkowski space string 
by the limit $\mu \rar 0$.  Yet, in the context of 
string theory, and finite temperature string theory in particular, this limit 
has to be taken with care; as we have seen here, the nature of the Hagedorn 
temperature has a nontrivial dependence on $\mu$.

 Naively, large scale
modifications to space-time do not affect ultraviolet physics. But in the 
example studied here, as
 in many other
string theoretic situations, 
this reasoning fails; perhaps
 this could have been expected here 
from the fact that the scalar curvature diverges at $z=0$. Certainly
within string theory, modular invariance implies that high energy
physics is 
dependent on
 the two-dimensional worldsheet Casimir
energy, whereas one would have expected the Casimir energy to be 
more relevant for light string excitations. Indeed, one way to interpret the
Hagedorn temperature in string theory is as the appearance of a
massless winding mode in the timelike direction.\footnote{From this
  viewpoint, the results of \cite{Cvetic:2002hi}, which studies
  T-duality along the timelike Killing vector, may be interesting.}  
The fact that strings
simultaneously know about the large and small scale structure means
that only in very controlled situations the high energy behavior is
that of the Minkowski string. The calculations here indicate that 
in situations where the 2D
Casimir energy is affected by the background this is not the case.

The dual description, using the BMN limit of $\cN=4$ SYM, adds 
another  
interesting layer to the analysis as the dual theory exists for parameters 
which appear to be forbidden on (ideal gas) 
string theory arguments.\footnote{The
  conjecture that the Hagedorn behaviour should signal a phase
  transition, which we partly based on this observation, 
has been found true \cite{blt}; see note added in proof.}  The 
simplest explanation is that the duality, at least in the form used here, is 
breaking down. There is also the possibility that the dual theory is telling us 
that somehow --- notwithstanding the results we have found --- 
the Hagedorn temperature 
on the pp-wave string side is not strictly limiting and can be
crossed.  This is 
tantalizing because
for string theories such as the IIB Minkowski string in which the
Hagedorn divergence is associated with a phase transition, the identification 
of the degrees of
freedom in the Hagedorn phase 
is one of the
important outstanding questions \cite{Atick:1988si} (see
\cite{Barbon:2001di} for the most recent progress). 
Perhaps, with further study beyond the ideal gas approximation, the existence
of a dual description for pp-wave string theory may give insight into this 
critical issue.

We conclude with a question. 
The traditional interpretation of the Hagedorn temperature, by
analogy with QCD, is a confinement/deconfinement transition. Such a
transition is known to occur for large $N$ 
$\cN=4$ SYM on $S_3$ when the
temperature $T$ is of order one in units of the $S_3$ radius
\cite{wittenads}, although 
it is unclear whether this transition is related to
Hagedorn behaviour of the AdS string (see, however, \cite{sundborg}). 
It would be interesting to see if the results in
\cite{wittenads} can be extended to include finite $R$-charge 
density as
well. Perhaps this might shed more light on the large $R$-charge 
density - finite temperature behaviour of $\cN=4$ SYM.

\bigskip
{\bf Note added in proof:}
As this article was being prepared for publication, a preprint by
Brower, Lowe and Tan appeared which improved 
the analysis of the asymptotic
behaviour of the partition function. 
For the value of the Hagedorn temperature, their
observations have no effect. However, the nature of the transition is
affected. Brower, Lowe and Tan's improved treatment of 
the $\theta \sim \tau_1$
Lagrange multiplier integral found that it
contributes an extra factor of $\sqrt{\tau_2}$ to the remaining 
$\tau_2$ integral. Combined with a Jacobian factor
$\tau_2$ from changing to $(\tau_2,\th=\tau_1/\tau_2)$ coordinates
this means that the measure in the final $\tau_2$ integral equals
$d\tau_2/\sqrt{\tau_2}$ rather than $d\tau_2/\tau_2^2$. This implies
that free energy is finite at $T_H$ and that the Hagedorn bound 
may demarcate a phase transition for pp-wave
strings. Similarly 
Brower, Lowe and Tan's results indicate that closed strings
compactified on $T^p$ have 
a non-limiting Hagedorn temperature for $p <9$.
 Our conjecture that the Hagedorn temperature should demarcate a phase
 transition is thereby borne out, and this opens up a new avenue to
 explore both the Hagedorn phase of string theory and the connection
 between pp-wave strings and $\cN=4$ super-Yang-Mills at finite
 chemical potential. We are grateful to 
R. Brower, D. Lowe, and C. Tan for comments and discussions.

\bigskip

{\bf Acknowledgments:}
We thank Maulik Parikh for collaboration in the early stages of this
work and many extensive and intensive discussions. We also thank 
David Berman, Jan de Boer, Chong-Sun Chu, Clifford Johnson, Dan Kabat, 
Al Mueller, Hiroshi Ooguri, 
Dam Thanh Son, Matt Strassler, and Bo Sundborg for
fruitful discussions and suggestions. KS and GS would like to thank
the organizers and participants of the Amsterdam Summer workshop on
String Theory and BG and GS would like to thank the
Aspen Center for Physics for providing stimulating environments to pursue this research. 
ISCAP gratefully acknowledges the
generous support of the Ohrstrom foundation. 
KS and BG gratefully acknowledge 
support from the DOE grant DE-FG-02-92ER40699. 
GS gratefully acknowledges
support from 
the DOE grant EY-76-02-3071, and the University of
Pennsylvania SAS Dean's funds. 

\appendix
\section{Subleading divergence of the Building Blocks}
\setcounter{equation}{0}

We seek the coefficient of the $\ln(\tau_2)$ divergence in eq. (\ref{eq:55})
\begin{eqnarray}
  \label{eq:app:1}
  -2 \sum_{k,\ell=1}^{\infty} e^{2\pi i (\ell \alpha -k\del)}
\frac{\tilde{\mu} K_1(\frac{\tilde{2\pi\mu}}{\tau_2} 
  \sqrt{(\ell\tau_2)^2+k^2})}
{\sqrt{(\ell\tau_2)^2+k^2}} ~+{\rm c.c.} \stackrel{\tau_2 \rar 0}{=}
c(\tilde{\mu},\alp,\del)\ln(\tau_2) + \ldots
\end{eqnarray}
in the limit $\tau_2 \rightarrow 0$ for $\tilde{\mu}=0$. Compared to
eq. (\ref{eq:55}) in the text, we have added
the complex conjugate contribution to the RHS; 
recall from eq. (\ref{eq:36}) and
eq. (\ref{eq:17}) that the building
blocks $\Theta_{\alp,\del}(\tau,\bar{\tau},\m)$ are real. This double
sum behaves differently in this limit for $\alpha \neq 0$ or $\del
\neq 0$ and $\alpha,\del=0$. The symmetry in the sums over $\ell$
and $k$ allows us to concentrate only on the case where either
$\alpha \neq 0$ or $\del \neq 0$. We will choose the former.

Noting that the expression can be rewritten so that the 
summation parameters,
$\ell$ and $k$, always multiply $\tilde{\mu}$, we may approximate for
$\tilde{\mu}$ small the
double 
sum by a double integral. This has to be done with care since the
summation is from $1$ to infinity. The resulting 
double integral 
equals
\begin{eqnarray}
 -2 \int_{\tilde{\mu}}^{\infty} dk \int_{\tilde{\mu}\tau_2}^{\infty}
  d\ell \, (e^{2\pi i \frac{\alpha}{\tilde{\mu}\tau_2}\ell}+e^{2\pi i \frac{\alpha}{\tilde{\mu}\tau_2}\ell} )
\frac{K_1(\frac{2\pi}{\tau_2} \sqrt{\ell^2+k^2})}
{\tau_2 \sqrt{\ell^2+k^2}}~.
\end{eqnarray}
We are integrating over the first quadrant of the 
$k,~\ell$ plane except for a rectangle
centered on the origin with area $4\tilde{\mu}^2\tau_2$. 
As $\mu$ is very small,
we approximate the lower boundary by a circle of the same area, and
change to polar coordinates. We obtain
\begin{eqnarray}
  \label{eq:app:3}
 -2 \int_{\tilde{\mu}\sqrt{\frac{2\tau_2}{\pi}}}^{\infty} rdr \frac{K_1(\frac{2\pi
  r}{\tau_2})}{\tau_2 r} \int_0^{\pi/2} d\theta \, (e^{2\pi i
  \frac{r\alpha \cos\theta}{\tilde{\mu}\tau_2}}+e^{-2\pi i
  \frac{r\alpha \cos\theta}{\tilde{\mu}\tau_2}})~.
\end{eqnarray}
The angular integral yields a Bessel function of the first kind, and
after a rescaling of $r$ one finds
\begin{eqnarray}
\label{eq:app:33}
-\int_{\tilde{\mu}\sqrt{\frac{8\pi}{\tau_2}}}^{\infty} dr \, K_1(r)
 J_0(\frac{r \alpha}{\tilde{\mu}})~. 
\end{eqnarray}
The integral (\ref{eq:app:33}) is convergent for $\tilde{\mu} =0$ 
at the upper
boundary, but not at the lower boundary. The
result depends on whether $\alpha$ does or does not vanish: 
the limits $\alpha \rar 0$, $\tilde{\mu} \rar 0$ do not commute.

If $\alpha \neq 0$, we may approximate the Bessel
function $J_0(r\alp/\tilde{\mu})$ for $\tilde{\mu} =0$ with its asymptote at infinity
\begin{eqnarray}
  \label{eq:app:15}
  x \rightarrow \infty~:~~~~~J_0(x) \rightarrow \sqrt{\frac{2}{\pi x}} \cos(x-\pi/4)~,
\end{eqnarray}
and obtain
\begin{eqnarray}
  \label{eq:app:311}
  -\int_{\tilde{\mu}\sqrt{\frac{8\pi}{\tau_2}}}^{\infty} dr \, K_1(r) J_0(\frac{r \alpha}{\tilde{\mu}})
 = -\int_{\tilde{\mu}\sqrt{\frac{8\pi}{\tau_2}}}^{\infty} dr \, K_1(r)
 \sqrt{\frac{2\tilde{\mu}}{\pi r \alpha }} \cos(\frac{r\alpha}{\tilde{\mu}}-\pi/4)~.
\end{eqnarray}
Substiting at the lower boundary the asymptote of 
$K_1(x)$ near $x=0$,
\begin{eqnarray}
  \label{eq:app:4}
  x \rightarrow 0~&:&~~~~~~K_1(x) \sim \frac{1}{x} ~,
\end{eqnarray}
one finds that the behaviour of the integral near the lower
boundary equals
\begin{eqnarray}
  \label{eq:app:18}
  -\int_{\tilde{\mu}\sqrt{\frac{8\pi}{\tau_2}}}^{\Lambda} dr \, \frac{1}{r}
 \sqrt{\frac{2\tilde{\mu}}{\pi r \alpha }} \cos(\frac{r\alpha}{\tilde{\mu}}-\pi/4)
 &=&-\sqrt{\frac{2\tilde{\mu}}{\pi\alpha}} \cos(\pi/4) \left[
 -\frac{2}{\sqrt{r}}\right]^{\Lambda}_{ \tilde{\mu}\sqrt{\frac{8\pi}{\tau_2}}} \non
&\simeq&  \sqrt{\frac{4}{\pi\alpha}}
 \cos(\pi/4)\left(\frac{\tau_2}{8\pi}\right)^{1/4} +f(\Lambda) +{{\cal O}}(\tilde{\mu})~.
\end{eqnarray}
The small $\tilde{\mu}$ behaviour is finite, and there is no
logarithmic divergence for $\tau_2 \rar 0$. Instead the integral
vanishes as a quarter power. Note,
however, that this power has a multiplicative constant which 
is inversely proportional to $\alpha$. The answer therefore 
only holds if $\alpha \neq 0$.

For $\alpha =0$ we should approximate the Bessel function $J_0(x)$
with its asymptote near zero rather than infinity. This is just unity.
Again approximating $K_1(x) \rar 1/x $ by its small $x$ asymptote we now find logarithmic
behaviour at the lower boundary 
\begin{eqnarray}
  \label{eq:app:5}
 -\int_{\tilde{\mu}\sqrt{\frac{8\pi}{\tau_2}}}^{\Lambda} dr \,
  \frac{1}{r} &=& 
  \ln(\tilde{\mu}\sqrt{\frac{8\pi}{\tau_2}})+ {{\cal O}}(\tilde{\mu}) +f(\Lam)
  \non
 &=&- \frac{1}{2} \ln \tau_2 +\ldots
\end{eqnarray}
We read off the expected result $c(\tilde{\mu}=0)=-\frac{1}{2}$.

Hence the building
blocks $\Theta_{\alpha,\del}(\tau,\bar{\tau},\m=0)$ have a subleading power law divergence for $\tau_2 \rar 0$, 
\begin{eqnarray}
  \label{eq:app:12}
  \The_{\alp,\del}(\tau,\bar{\tau};0) \sim
  \tau_2^{c(\alp,\del)} e^{\frac{\xi(\tilde{\mu}=0)}{\tau_2}}~, 
\end{eqnarray}
with
$c(\alpha \neq 0 ~{\rm or}~ \del \neq 0) =0$; 
$c(\alpha,\del=0)=-\hlf$, as was known by other methods.

\end{document}